\documentclass[prc,twocolumn,showpacs,preprintnumbers,amsmath,amssymb,superscriptaddress,floatfix,nofootinbib]{revtex4}
\usepackage{graphicx}
\usepackage{amsmath}
\usepackage{amsfonts}
\usepackage{amssymb}
\usepackage{color}

\begin{document}
\title{The $\gamma p\to n a^+_2(1320) \to n \rho^0 \pi^+$ reactions within an effective Lagrangian approach}

\author{Yin Huang}
\email{huangy2011@lzu.cn} \affiliation{Research Center for Hadron
and CSR Physics, Institute of Modern Physics of CAS and Lanzhou
University, Lanzhou 730000, China}  \affiliation{Institute of modern
physics, Chinese Academy of Sciences, Lanzhou 730000, China}
\affiliation{School of Nuclear Science and Technology, Lanzhou
University, Lanzhou 730000, China}
\author{Ju-Jun Xie} \email{xiejujun@impcas.ac.cn}
\affiliation{Research Center for Hadron and CSR Physics, Institute
of Modern Physics of CAS and Lanzhou University, Lanzhou 730000,
China} \affiliation{Institute of modern physics, Chinese Academy of
Sciences, Lanzhou 730000, China} \affiliation{State Key Laboratory
of Theoretical Physics, Institute of Theoretical Physics, Chinese
Academy of Sciences}
\author{Xu-Rong Chen}
\affiliation{Research Center for Hadron and CSR Physics, Institute
of Modern Physics of CAS and Lanzhou University, Lanzhou 730000,
China} \affiliation{Institute of modern physics, Chinese Academy of
Sciences, Lanzhou 730000, China}
\author{Jun He}
\affiliation{Research Center for Hadron and CSR Physics, Institute
of Modern Physics of CAS and Lanzhou University, Lanzhou 730000,
China}  \affiliation{Institute of modern physics, Chinese Academy of
Sciences, Lanzhou 730000, China} \affiliation{State Key Laboratory
of Theoretical Physics, Institute of Theoretical Physics, Chinese
Academy of Sciences}
\author{Hong-Fei Zhang}
\affiliation{Institute of modern physics, Chinese Academy of
Sciences, Lanzhou 730000, China}  \affiliation{School of Nuclear
Science and Technology, Lanzhou University, Lanzhou 730000, China}

\date{\today}

\begin{abstract}

We investigate the $a_2(1320)$ meson photon production in the
$\gamma p \to n a^+_2(1320)$ and $\gamma p \to n \rho^0 \pi^+$
reactions within the effective Lagrangian method. For $\gamma p \to
n a^+_2(1320)$ reaction, by considering the contributions only from
the $t-$channel $\pi^+$ exchange, we get a fairly good description
of the current experimental total cross section data. We also
studied the $\gamma p \to n \rho^0 \pi^+$ reaction, which mainly
contribute to the $\gamma p \to n \pi^+ \pi^+ \pi^-$ reaction. The
latter reaction has measured by the CLAS Collaboration. The total
cross sections, invariant mass distribution, and the Dalitz Plot of
$\gamma p \to n \rho^0 \pi^+$ reaction are shown, which can be
tested by future experiments.

\end{abstract}

\maketitle
\section{INTRODUCTION}

In the traditional constituent quark models (CQM), mesons are
described as quark-anti-quark ($q\bar{q}$) states. This picture
could explain successfully the properties of the ground states of
the flavor SU(3) vector meson nonet. However, there are many meson
(or meson-like) states could not be explained as $q\bar{q}$ states.
For example, in the low energy scalar sector, the $\sigma(500)$,
$a_0(980)$ and $f_0(980)$~\cite{pdg2012}, which are proposed as the
meson-meson dynamically generated
states~\cite{isgurprl48,ollerppnp45}. Besides, those meson states,
with spin-parity-charge parity $J^{PC}=0^{--}$, $0^{+-}$, $1^{-+}$,
and $2^{+-}$, \emph{etc}, would be also existence, but they cannot
be obtained by $q\bar{q}$ pairs within the CQM. These states are
known as "exotics" and observation of them has been of great
interest as it would be clear evidence for mesons beyond the
classical CQM picture.

Indeed, on the experimental side, searching for the scalar-isoscalar
mesons have been carried out in the $p\bar{p}$
annihilation~\cite{amslerpmp70}, $pp$ interaction at high
energies~\cite{Antinori:1995wz,Barberis:1997ve,Barberis:1999an},
$\pi N$
reactions~\cite{Alde:1988bv,Beladidze:1993km,Aoyagi:1993kn,Thompson:1997bs},
$J/\psi$ radiative decays~\cite{Bugg:1995jq,Bai:1996dc}, and photon
production processes~\cite{Szczepaniak:2001qz}. The results show
that in the $1-2$ GeV mass range there are several meson states that
do not agree with the predictions of the CQM, and some of these
states may have a significant non-$q\bar{q}$ component. As in
Ref.~\cite{Alde:1988bv}, an exotic meson, $J^{PC}=1^{-+}$, $I^G =
1^-$, with mass $1406 \pm 20$ MeV and width $180 \pm 30$ MeV, has
been observed in the study of the exclusive reaction $\pi^- p \to
\pi^0 \eta n$ at $100$ GeV. Ten years later, the E852 Collaboration
reported another exotic resonant signal in the $\pi^- p \to \pi^-
\eta p$ reaction at $18$ GeV~\cite{Thompson:1997bs}, the
corresponding mass and width are $1370 \pm 16 ^{+50}_{-30}$ MeV and
$385 \pm 40 ^{+65} _{-105}$ MeV, respectively. Furthermore, lattice
and model calculations indicate that the lightest exotic meson
should have $J^{PC}=1^{-+}$ and mass below $2$
GeV~\cite{bernardprd56,isgurprd31}.

Recently, the CLAS Collaboration at Jefferson Lab has contributed in
looking for the exotic mesons in the $\pi^+ \pi^+ \pi^-$ system
photoproduced by the charge exchange reaction $\gamma p \to \pi^+
\pi^+ \pi^- n$~\cite{Nozar:2008aa}. However, the partial wave
analysis shows that the most contributions to the $\pi^+ \pi^+
\pi^-$ production are from the tensor mesons $a_2(1320)$
[$I^G(J^{PC}) = 1^-(2^{++})$] and $\pi_2(1670)$. There is no
evidence for the production of the exotic states $a_1(1260)$ and
$\pi_1(1600)$ at expected levels.

Since the $\rho \pi$ channel is the main decay channel of the
$a_2(1320)$ ($\equiv a_2$) meson, the $\gamma p \to n a^+_2(1320)$
reaction has been studied from the $\gamma p \to n (\rho \pi)^+$
reaction by several
experiments~\cite{CambridgeBubbleChamberGroup:1968zz,Nozar:2008aa,Ballam:1969jv,Eisenberg:1969kk,Ballam:1971wq}.
On the theoretical side, a phenomenological analysis, for studying
the production mechanism of the exotic states, has been done in
Ref.~\cite{Szczepaniak:2001qz} with the vector meson dominance (VMD)
model. They found that in the photo-production the exotic
$\pi_1(1600)$ and $a_2(1320)$ meson production should be comparable,
which is disagreement with the recent precise experimental
measurements~\cite{Nozar:2008aa}. Thus, more theoretical studies on
this issue are needed.

In the present work, with the new experimental results from CLAS
Collaboration~\cite{Nozar:2008aa}, we study the $\gamma p\to n
a^+_2(1320) \to n \rho^0 \pi^+$ reactions by using the effective
Lagrangian approach and the isobar model. We consider the
contributions from the $t$-channel $\pi^+$ exchange for the $\gamma
p\to n a^+_2(1320)$ reaction. For the low energy of the $\gamma p
\to n \rho^0 \pi^+$ reaction, we pay especially attention on the
role of the $a_2(1320)$ meson.

This paper is organized as follows. In Sect.~\ref{Formalism}, we
present the formalism and ingredients necessary for our
calculations. The numerical results are also shown. A short summary
is given in the last section.

\section{Formalism and numerical results} \label{Formalism}

\begin{figure}[htbp]
\begin{center}
\includegraphics[scale=0.6]{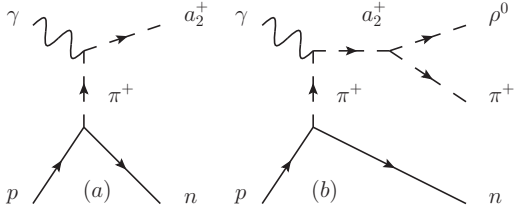}
\caption{Feynman diagrams for $\gamma p \to na^+_2$ and $\gamma p
\to n \pi^+ \rho^0$ reactions.} \label{feydiagram}
\end{center}
\end{figure}

The basic tree level Feynman diagrams, for the $\gamma p \to n
a^+_2(1320)$ [Fig.~\ref{feydiagram} (a)] and $\gamma p \to n \rho^0
\pi^+$ [Fig.~\ref{feydiagram} (b)] reactions, are shown in
Fig.~\ref{feydiagram}, where we consider only the $t-$channel
$\pi^+$ exchange process, while the $s-$channel and $u-$channel
processes are neglected since the information of those processes is
scarce and we expect these contributions to be small.

\subsection{$\gamma p \to n a^+_2$ reaction}

Firstly, we pay attention to the reaction $\gamma p \to n a^+_2$. To
compute the contributions from the $t-$channel $\pi^+$ exchange, as
shown in Fig.~\ref{feydiagram} (a), we use the effective interaction
Lagrangian densities as used in
Refs.~\cite{Bongardt:1979qf,Levy:1975fk,Babcock:1976hr,Uy:1983kj},
\begin{eqnarray}
{\cal L}_{a_2\pi\gamma} & = &  \frac{g_{a_2\pi\gamma}}{m_{\pi}^2}
\epsilon_{\mu \nu \rho \sigma} \partial^{\mu}
a_2^{\nu\delta}\partial^{\rho}A^{\sigma}\partial_{\delta}\phi_{\pi},
\label{la2pigama} \\
{\cal L}_{\pi NN} &=& -i g_{\pi
NN}\bar{N}\gamma_{5}\vec{\tau}\cdot{\vec{\pi}} N,
\end{eqnarray}
where $g^2_{\pi NN}/4\pi=14.4$, while the value of the coupling
constant $g_{a_2\pi \gamma}$ can be determined from the partial
decay width of $a_2 \to \pi \gamma$, which can be easily obtained
with Eq.~(\ref{la2pigama}),
\begin{equation}
\Gamma_{a_2 \to \pi \gamma}=\frac{g_{a_2\pi \gamma}^2}{40\pi
m^4_{\pi}}p^5_{\gamma}, \label{a2topigama}
\end{equation}
with
\begin{eqnarray}
p_{\gamma} = \frac{M^2-m^2_{\pi}}{2M},
\end{eqnarray}
where $M$ is mass of $a_2(1320)$ meson.

With mass ($M = 1318.3$ MeV, $m_{\pi} = 139.57$ MeV), total decay
width ($\Gamma_{a_2} = 107$ MeV), and decay branching ratio of $a_2
\to \pi \gamma$ [Br($a_2 \to \pi \gamma) = 2.68 \times 10^{-3}$ ],
from Eq.~(\ref{a2topigama}), we obtain $g_{a_2 \pi \gamma} = 1.08
\times 10^{-2}$.

As we are not dealing with point-like particles, we ought to
introduce the compositeness of the hadrons. This is usually achieved
by including form factors in the finite interaction vertexes. In the
present work, we adopt the following form factor as used in the Bonn
model~\cite{machleidt} for the exchanged $\pi^+$ meson in the
$t$-channel,
\begin{equation}
F_{\pi}(t)=\frac{\Lambda_{\pi}^2-m_{\pi}^2}{\Lambda_{\pi}^2-t},
\end{equation}
where $t = q^2$ with $q$ the four momentum of the exchanged $\pi$
meson, and $\Lambda_{\pi}$ is the cutoff parameter.

With the effective Lagrangian densities given above, we can easily
construct the invariant scattering amplitude for the $\gamma p \to n
a^+_2$ reaction,
\begin{eqnarray}
{\cal M} = \bar u(s_2,p_2)~{\cal A}_{\sigma, \nu \delta}
\varepsilon^{\sigma}(k_1,\lambda_1) T^{\nu \delta
*}(k_2,\lambda_2)~u(s_1,p_1),
\end{eqnarray}
where $s_2,p_2$ and $s_1,p_1$ denote the spin polarization variables
and the four-momenta of the outgoing neutron and the initial proton,
respectively, while $k_1,\lambda_1$ and $k_2,\lambda_2$ are the
four-momenta and spin polarization variables of the photon and
$a_2(1320)$ meson, respectively. The $\bar u(s_2,p_2)$ and
$u(s_1,p_1)$ are the Dirac spinors for the neutron and proton,
respectively, while the $\varepsilon_{\mu}(k_1,\lambda_1)$ and
$T^*_{\nu \rho}(k_2,\lambda_2)$ are the polarization vector and the
polarization tensor of the photon and $a_2(1320)$ meson,
respectively. The reduced ${\cal A}_{\sigma, \nu \delta}$ reads,
\begin{eqnarray}
{\cal A}_{\sigma, \nu \delta} &=& i \frac{g_{\pi NN}g_{a_2 \pi
\gamma}}{m^2_{\pi}} F^2_{\pi}(t) \frac{\epsilon_{\mu \nu \rho
\sigma} k^{\mu}_2 k^{\rho}_1 q_{\delta}}{ t - m_{\pi}^2}.
\end{eqnarray}

Then, the differential cross section for $\gamma p \to n a^+_2$ at
center of mass (c.m.) frame can be expressed as
\begin{equation}
{d\sigma \over d{\rm cos}\theta}={m_p m_n \over 8\pi s}{
|\vec{k_2}^{\text{c.m.}}| \over |\vec{k_1}^{\text{c.m.}}|} \left (
{1\over 4}\sum_{s_1,s_2,\lambda_1,\lambda_2}{|\cal M|}^2 \right ),
\label{eq:dcs}
\end{equation}
where $\theta$ denotes the angle of the outgoing $a^+_2$ meson
relative to beam direction in the $\rm c.m.$ frame,
$\vec{k_1}^\text{c.m.}$ and $\vec{k_2}^\text{c.m.}$ are the
3-momentum of the initial photon and the final $a^+_2$ meson in the
$\rm c.m.$ frame, respectively.

The sum over polarizations, in Eq.~(\ref{eq:dcs}), can be easily
done thanks to
\begin{eqnarray}
\sum_{\lambda} \varepsilon^{\mu} (k_1,\lambda) \varepsilon^{\nu *}
(k_1,\lambda) = -g^{\mu\nu} ,
\end{eqnarray}
for the photon, and
\begin{eqnarray}
&& \sum_{\lambda} T_{\mu\nu} (k_2,\lambda) T_{\mu^{'}\nu^{'}}^*
(k_2,\lambda) = P_{\mu \nu \mu' \nu'}  \nonumber \\
&&  = \frac{1}{2}(\widetilde{g}_{\mu\mu^{'}}
\widetilde{g}_{\nu\nu^{'}} + \widetilde{g}_{\mu \nu^{'}}
\widetilde{g}_{\nu\mu^{'}} )  -
\frac{1}{3}\widetilde{g}_{\mu\nu}\widetilde{g}_{\mu^{'}\nu^{'}},
\label{a2spinsum}
\end{eqnarray}
for the tensor $a_2(1320)$ meson, where $\widetilde{g}_{\mu\nu}=
-g^{\mu\nu} + \frac{k_2^{\mu}k_2^{\nu}}{M^2}$.

\begin{figure}[htbp]
\begin{center}
\includegraphics[scale=0.4]{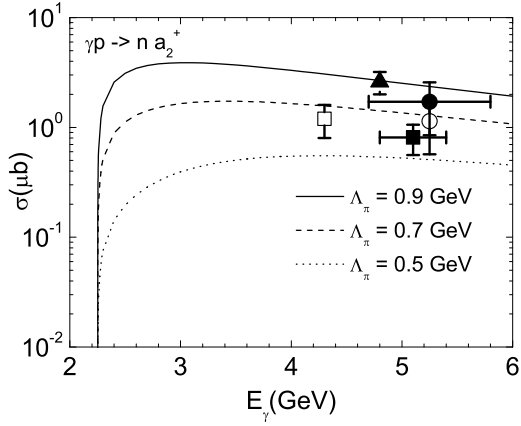}
\caption{The total cross sections of $\gamma p \to n a^+_2$ reaction
as function of photon energy $E_{\gamma}$. The experimental data are
taken from Ref.~\cite{Nozar:2008aa} (square),
Ref.~\cite{Ballam:1969jv} (cycle), Ref.~\cite{Eisenberg:1969kk}
(open square), Ref.~\cite{Ballam:1971wq} (dot), and
Ref.~\cite{condoprd48} (triangle). } \label{Fig:tcsna2}
\end{center}
\end{figure}

With the ingredients shown above, we can easily calculate the total
cross sections for $\gamma p \to n a^+_2$ reaction, which is shown
in Fig.~\ref{Fig:tcsna2}, where the solid, dashed, and dotted lines
are obtained with cut off parameter $\Lambda_{\pi} = 0.9$, $0.7$,
and $0.5$ GeV, respectively. The experimental data from
Refs.~\cite{Nozar:2008aa,Ballam:1969jv,Eisenberg:1969kk,Ballam:1971wq,condoprd48}
are also shown for comparing. We see that our theoretical results,
which is obtained including only the contributions from the
$t-$channel $\pi^+$ exchange, can give a reasonable description of
the current experimental data.

\subsection{$\gamma p \to n \rho^0 \pi^+$ reaction}

Next, we pay attention to the $\gamma p \to n \rho^0 \pi^+$
reaction, which mainly contribute to the $\gamma p \to n \pi^+ \pi^+
\pi^-$ reaction.

\begin{figure}[htbp]
\begin{center}
\includegraphics[scale=0.6]{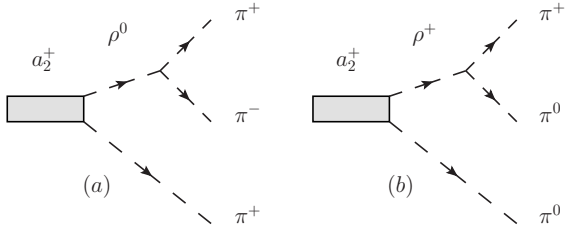}
\caption{Feynman diagrams for $a_2\rightarrow{}3\pi$ reactions.} \label{Fig:feydiagram}
\end{center}
\end{figure}

In the present case, we need also the interaction Lagrangian density
for the $a_2 \pi \rho$ vertex,
\begin{eqnarray}
{\cal L}_{a_2\pi\rho} & =& \frac{g_{a_2\pi\rho}}{m_{\pi}^2}
\epsilon_{\mu\nu\eta\sigma}\partial^{\mu}a_2^{\nu\delta}\partial^{\eta}\rho^{\sigma}\partial_{\delta}\phi_{\pi},
\label{la2pirho}
\end{eqnarray}
where the value of the coupling constant $g_{a_2 \pi \rho }$ can be
determined from the $a_2 \to \rho \pi \to \pi \pi \pi$ decays. The
decay processes are shown in Fig.~\ref{Fig:feydiagram}.

We follow the formalism as in Ref.~\cite{xiewilkin} for the case of
$N^*(1535)N\rho$ coupling. The $a_2 \to \pi\pi\pi$ partial decay
width is related to the decay amplitude through
\begin{eqnarray}
&& d\Gamma_{a_2 \to \rho \pi \to \pi \pi \pi}  = \frac{1}{2M}
\frac{3}{5}|\mathcal{M}_{a_2^{+} \to \rho^{0} \pi^{+} \to \pi^{+}
\pi^{-} \pi^{0}}|^2 \times \nonumber \\
&& \frac{1}{(2\pi)^5}\frac{d^3p_1d^3p_2d^3p_3}{8E_1E_2E_3}
\delta^4(M\!-\!p_1\!-\!p_2\!-\!p_3)\,, \label{pipidecay}
\end{eqnarray}
where $p_1, p_2, p_3$ and $E_1, E_2, E_3$ are the momenta and
energies of the final mesons. The decay amplitude
$\mathcal{M}_{a_2^{+} \to \rho^{0} \pi^{+} \to \pi^{+} \pi^{-}
\pi^{0}}$ is,
\begin{eqnarray}
\mathcal{M}_{a_2^{+} \to \rho^{0} \pi^{+} \to \pi^{+} \pi^{-}
\pi^{0}} &=& \frac{g_{\rho\pi\pi} g_{a_2\pi\rho}}{m_{\pi}^2}
\epsilon_{\mu\nu\eta\sigma}p_{a_2}^{\mu}T^{\nu\delta}p_{\rho}^{\eta}p_{1 \delta} \times \nonumber \\
&& G^{\sigma\beta}(p_{\rho})(p_2-p_3)_{\beta}F_{\rho}(p^2_{\rho}),
\end{eqnarray}
where $p_{a_2} = p_1 + p_2 + p_3$ and $p_{\rho} = p_2 + p_3$ are the
four momenta of the $a_2$ meson and $\rho$ meson, respectively. The
$F_{\rho}(p^2_{\rho})$ and $G_{\sigma\beta}(p_{\rho})$ are the form
factor and the propagator for the $\rho$-meson, with the forms as in
Ref.~\cite{xiewilkin},
\begin{eqnarray}
F(p^2_{\rho}) &=&
\frac{\Lambda^2_{\rho}}{\Lambda^2_{\rho}+|p^2_{\rho}-m^2_{\rho}|}\,, \label{sff}\\
G^{\sigma\beta}(p_{\rho}) &=&
-i\frac{g^{\sigma\beta}-p_{\rho}^{\sigma}p_{\rho}^{\beta}/p_{\rho}^2}{p_{\rho}^2-m_{\rho}^2+im_{\rho}\Gamma_{\rho}},
\end{eqnarray}
with a cut-off parameter $\Lambda_{\rho}$ and the total decay width
of $\rho$ meson, $\Gamma_{\rho} =150$ MeV .

\begin{figure}[htbp]
\begin{center}
\includegraphics[scale=0.4]{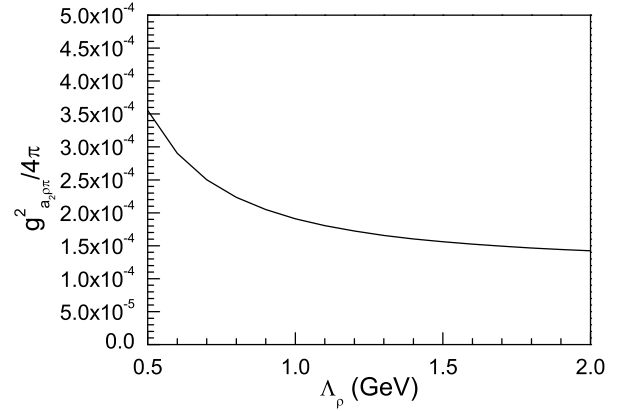}
\caption{Coupling constant $g^2_{a_2\rho\pi}/4\pi$ \emph{versus} the
cut-off parameter $\Lambda_{\rho}$.}\label{ga2rhopi}
\end{center}
\end{figure}

With the values of $g^2_{\rho \pi \pi}/4\pi = 2.91$ and the partial
decay width $\Gamma_{a_2 \to \rho \pi \to \pi \pi\pi} = 75$
MeV~\cite{pdg2012}, we can get the coupling constant
$g^2_{a_2\rho\pi}/4\pi$ as a function of the cut off parameter
$\Lambda_{\rho}$ as shown in Fig.~\ref{ga2rhopi}. In the present
calculation, we will take $\Lambda_{\rho} = 1.0$ GeV which leads to
$g^2_{a_2\rho\pi}/4\pi = 1.9 \times 10^{-4}$.

Turning now to the $\gamma p \to n \rho^0 \pi^+$ reaction, by using
the formalism and ingredients given above, the calculations of the
differential and total cross sections for this reaction are
straightforward,
\begin{eqnarray}
&& d\sigma (\gamma p \to n\rho^0
\pi^+) = \frac{1}{8 E_{\gamma}} \sum_{s_1, s_2} \sum_{s_3, s_4} |{\cal M}|^2 \times \nonumber \\
&& \frac{m_n d^{3} p_{3}}{E_{3}} \frac{d^{3} p_4}{2 E_4} \frac{d^{3}
p_5}{2 E_5} \delta^4 (p_{1}+p_{2}-p_{3}-p_{4}-p_5),
\label{nrhopidcs}
\end{eqnarray}
where $p_i$ ($i=1, 2, 3, 4, 5$) is the four momentum for photon,
proton, neutron, $\rho^0$ meson, and $\pi^+$ meson, and $s_i$ ($i=1,
2, 3, 4$) is the spin polarization variables of photon, proton,
neutron and $\rho^0$ meson, while $E_{\gamma}$ is the photon energy
at the Lab frame. The scattering amplitude ${\cal M}$ is,
\begin{eqnarray}
&&{\cal M}= i
\frac{\sqrt{2}g_{a_2\pi\rho}g_{\gamma\pi{}a_2}g_{\pi{}N
N}}{m_{\pi}^4}\cal{A}_{\pi^+},
\end{eqnarray}
with
\begin{eqnarray}
\cal{A_{\pi^+}} &=& \bar{u}_{n}(p_3)\gamma_{5}u_{p}(p_{2})
\frac{F_{\pi}^2(t)}{t-m_{\pi}^2}
\epsilon_{\mu\nu\rho\sigma}q_{a_2}^{\mu}p_{1}^{\rho}\epsilon^{\sigma}(p_1)(p_2-p_3)_{\delta}
\nonumber \\
&& \times
\frac{F_{a_2}(q_{a_2}^2)P^{\nu\delta\beta\omega}}{q_{a_2}^2-M^2+iM\Gamma_{a_2}}
\epsilon_{\alpha\beta\eta\lambda}q_{a_2}^{\alpha}p_{4}^{\eta}\epsilon^{\lambda
*}(p_4) p_{5\omega}, \label{apiplus}
\end{eqnarray}
where $F_{a_2}(q^2_{a_2})$ is the form factor for the off shell
$a_2(1320)$ meson, which has the form,
\begin{eqnarray}
F_{a_2}(q^2_{a_2}) = \frac{\Lambda^4_{a_2}}{\Lambda^4_{a_2} +
(q^2_{a_2} - M^2)^2},
\end{eqnarray}
with $q_{a_2}$ is the four momentum of the $a_2(1320)$ meson and the
cut off parameter $\Lambda_{a_2} =1.0$ GeV. On the other hand, the
$P_{\beta\omega\nu\delta}$, in Eq.~(\ref{apiplus}), has been defined
in Eq.~(\ref{a2spinsum}).

\begin{figure}[htbp]
\begin{center}
\includegraphics[scale=0.30]{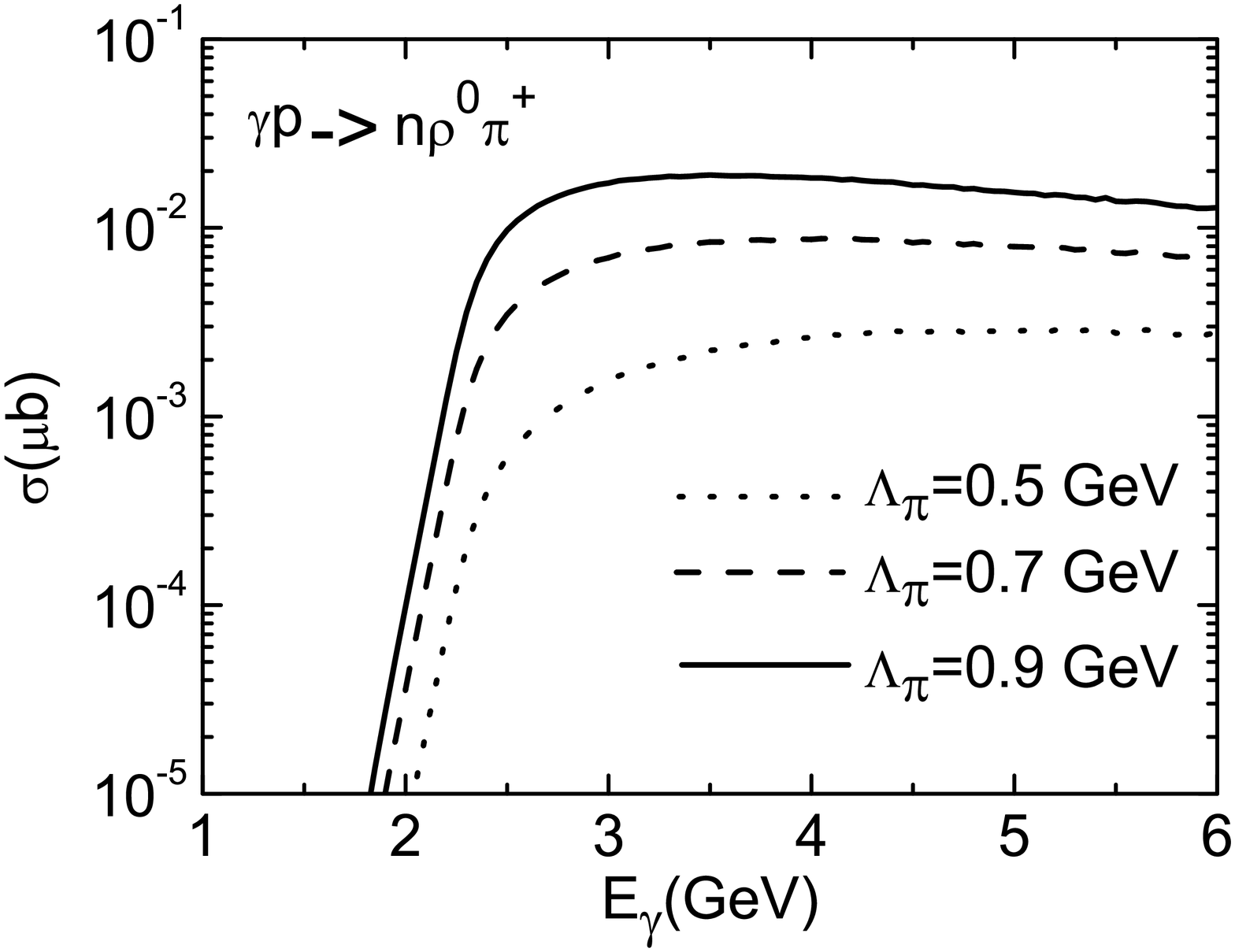}
\caption{The total cross section of $\gamma p \to n\rho^0 \pi^+$
reaction as function of photon energy $E_{\gamma}$ at the Lab
frame.} \label{Fig:tcsnrhopi}
\end{center}
\end{figure}

Then, the total cross section versus the beam energy $E_{\gamma}$ of
the photon for the $\gamma p \to n \rho^0 \pi^+$ reaction is
calculated by using a Monte Carlo multi-particle phase space
integration program. Our predictions, with $\Lambda_{\pi}= 0.5$,
$0.7$ and $0.9$ GeV, for the beam energies $E_{\gamma}$ from just
above the production threshold $1.36$ GeV to $6.0$ GeV are shown in
Fig.~\ref{Fig:tcsnrhopi} by dotted, dashed and solid curves,
respectively.

\begin{figure*}[htbp]
\begin{center}
\includegraphics[scale=0.35]{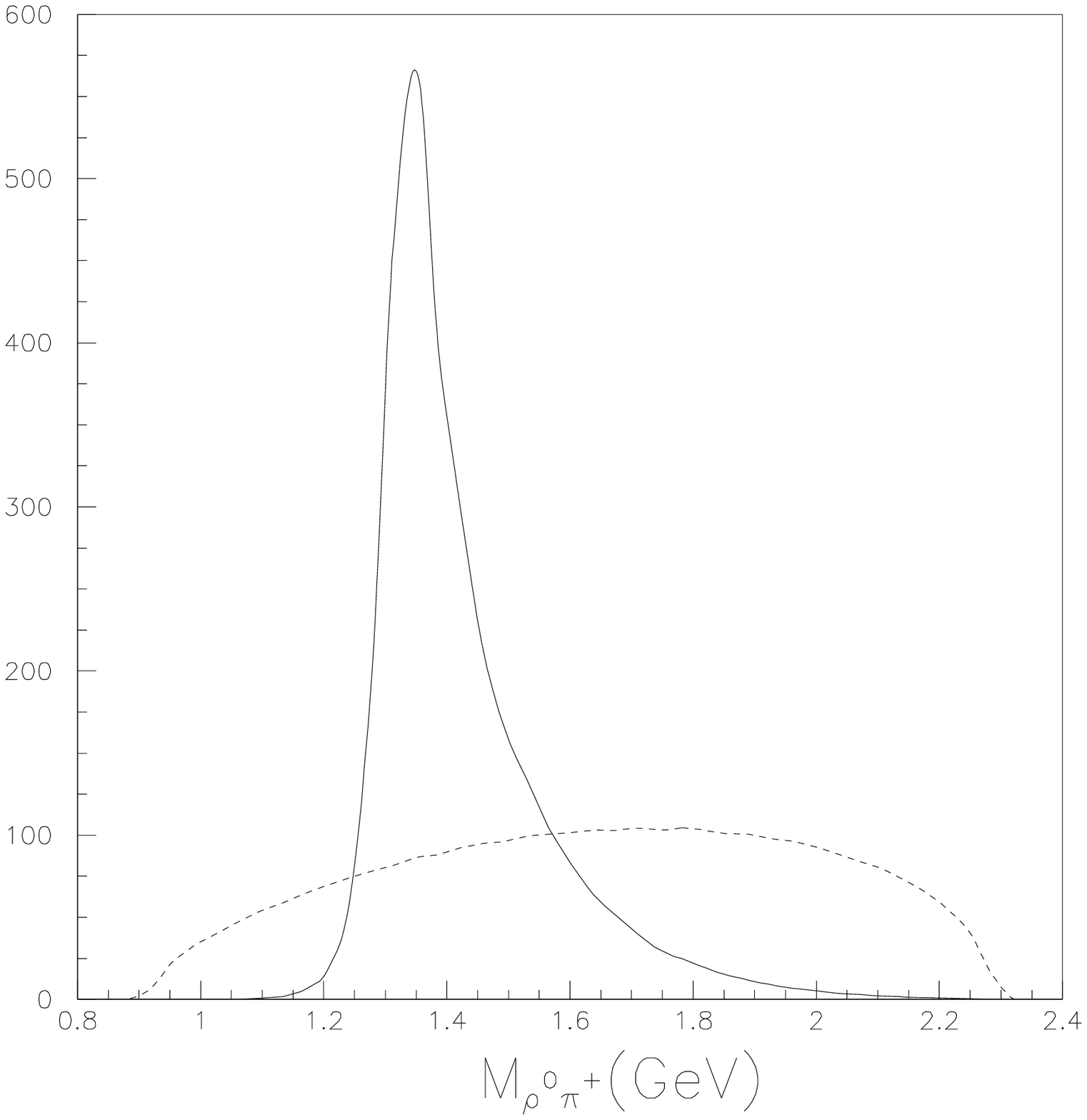}
\includegraphics[scale=0.35]{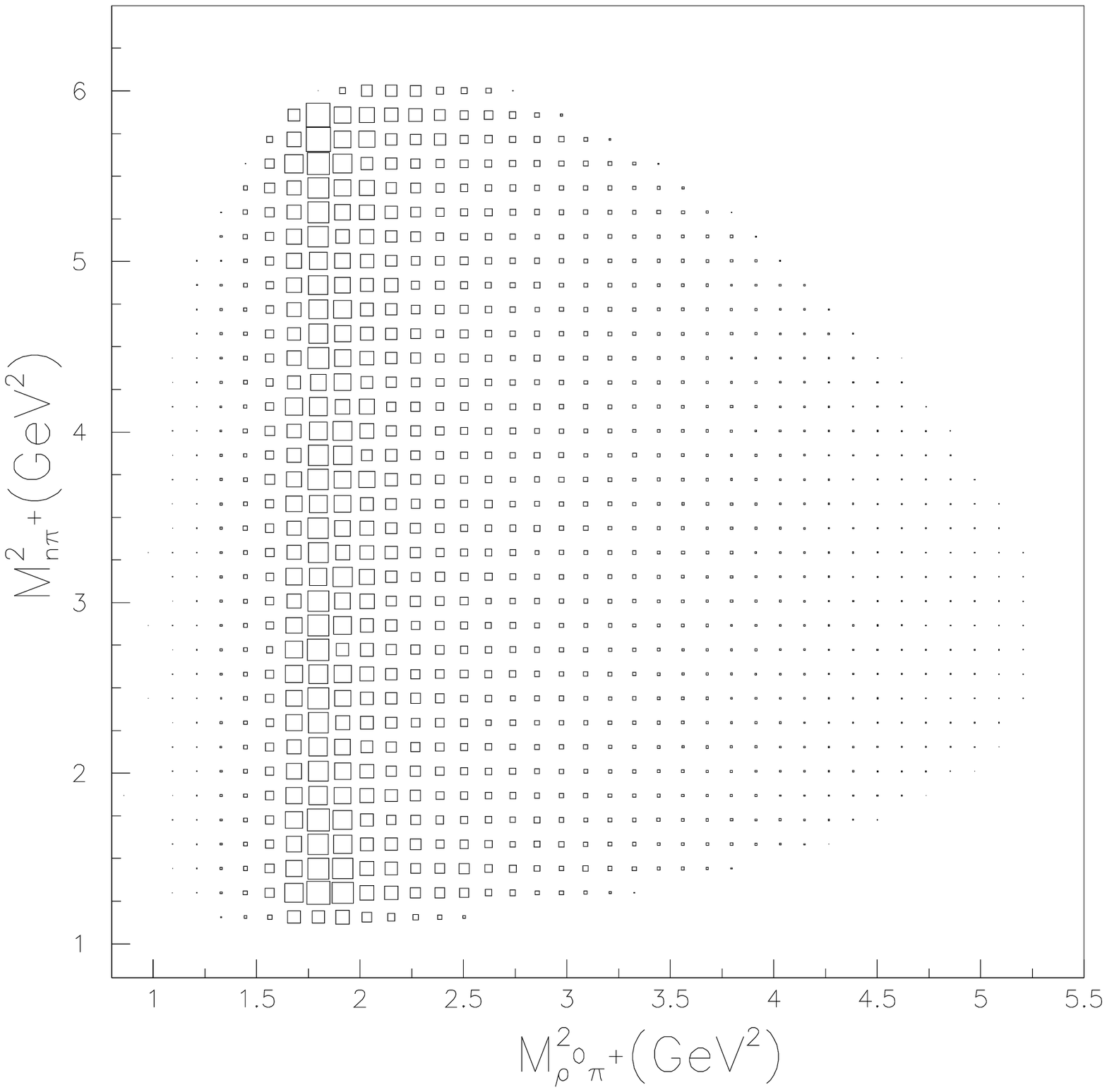}
\caption{The $\rho^0 \pi^+$ invariant mass spectrum (upper panel),
and the Dalitz Plot (lower panel) for the $\gamma p \to n \rho^0
\pi^+$ reaction at beam energy $E_{\gamma} = 5.1$ GeV. The dashed
lines are pure phase space distributions, while, the solid lines are
full results from our model.} \label{Fig:dp}
\end{center}
\end{figure*}

Furthermore, the corresponding $\rho^0 \pi^+$ invariant mass
spectrum, and the Dalitz Plot for the $\gamma p \to n \rho^0 \pi^+$
reaction at beam energy $E_{\gamma} = 5.1$ GeV, which is reachable
for the CLAS experiment~\cite{Nozar:2008aa}, are calculated and
shown in Fig.~\ref{Fig:dp}. The dashed lines are pure phase space
distributions, while, the solid lines are full results from our
model. From Fig.~\ref{Fig:dp}, we can see that there is a clear peak
in the $\rho^0 \pi^+$ invariant mass distribution, which is produced
by including the contribution from the $a_2(1320)$ meson. Those
theoretical predictions can be checked by the future experiments.

\section{Summary}

In this work, with the new experimental results from CLAS
Collaboration~\cite{Nozar:2008aa}, we perform a calculation of the
$a_2(1320)$ meson photon-production in the $\gamma p \to n
a^+_2(1320)$ and $\gamma p \to n \rho^0 \pi^+$ reactions within the
effective Lagrangian method and the isobar model. For the $\gamma p
\to n a^+_2(1320)$ reaction, by considering the contributions from
only the $t$-channel $\pi^+$ exchange, we get a fairly good
description of the current experimental total cross section data.

The recent results of the $\gamma p \to n \rho^0 \pi^+$ reaction
from CLAS Collaboration~\cite{Nozar:2008aa} show that the most
contributions to the $\pi^+ \pi^+ \pi^-$ production are from the
tensor mesons $a_2(1320)$ and $\pi_2(1670)$. So, basing on our
results of $\gamma p \to n a^+_2(1320)$ reaction, we have studied
the $\gamma p \to n \rho^0 \pi^+$ reaction, which mainly contribute
to the $\gamma p \to n \pi^+ \pi^+ \pi^-$ reaction. In this case, we
pay especially attention on the role of the $a_2(1320)$ meson. We
have calculated the total cross sections, invariant mass
distribution, and the Dalitz Plot for the $\gamma p \to n \rho^0
\pi^+$ reaction. Those theoretical predictions can be tested by the
future experiments.

\section*{Acknowledgments}

We would like to thank Dian-Yong Chen for useful discussions. This
work is partly supported by the National Natural Science Foundation
of China under Grants No. 11105126, No. 11275235, No. 11035006, No.
10905077, and the Knowledge Innovation Project of the Chinese
Academy of Sciences under Grant No.KJCX2-EW-N01.

\end{document}